# Depth resolved pencil beam radiography using AI – a proof of principle study

**Ida Häggström, PhD[a,b]; Lukas M. Carter, PhD[b] ; Thomas J. Fuchs, Prof, DSc[c]; Adam L. Kesner, PhD[b]**

[a] *Dept. of Radiology, Memorial Sloan Kettering Cancer Center,*
  *New York, USA*

[b] *Dept. of Medical Physics, Memorial Sloan Kettering Cancer Center,*
  *New York, USA*

[c] *Dept. of Pathology, Hasso Plattner Institute for Digital Health, Mount Sinai Medical School,*
  *New York, USA*

  *E-mail*: kesnera@mskcc.org

ABSTRACT: AIMS: Clinical radiographic imaging is seated upon the principle of differential keV photon transmission through an object. At clinical x-ray energies the scattering of photons causes signal noise and is utilized solely for transmission measurements. However, scatter – particularly Compton scatter, is characterizable. In this work we hypothesized that modern radiation sources and detectors paired with deep learning techniques can use scattered photon information constructively to resolve superimposed attenuators in planar x-ray imaging. METHODS: We simulated a monoenergetic x-ray imaging system consisting of a pencil beam x-ray source directed at an imaging target positioned in front of a high spatial- and energy-resolution detector array. The setup maximizes information capture of transmitted photons by measuring off-axis scatter location and energy. The signal was analyzed by a convolutional neural network, and a description of scattering material along the axis of the beam was derived. The system was virtually designed/tested using Monte Carlo processing of simple phantoms consisting of 10 pseudo-randomly stacked air/bone/water materials, and the network was trained by solving a classification problem. RESULTS: From our simulations we were able to resolve traversed material depth information to a high degree, within our simple imaging task. The average accuracy of the material identification along the beam was 0.91±0.01, with slightly higher accuracy towards the entrance/exit peripheral surfaces of the object. The average sensitivity and specificity was 0.91 and 0.95, respectively. CONCLUSIONS: Our work provides proof of principle that deep learning techniques can be used to analyze scattered photon patterns which can constructively contribute to the information content in radiography, here used to infer depth information in a traditional 2D planar setup. This principle, and our results, demonstrate that the information in Compton scattered photons may provide a basis for further development. The work was limited by simple testing scenarios and without yet integrating complexities or optimizations. The ability to scale performance to the clinic remains unexplored and requires further study.

# Contents

## 1. Introduction

With more than a century of development, various concepts and architectures for clinical x-ray imaging have been developed and refined to form the technology that supports the field of Radiology, a pillar of modern medicine. Radiography, computed tomography, fluoroscopy, and other x-ray technologies support numerous areas of health care, research, and industry.

Current clinical x-ray technologies are almost universally built upon the principle of differential transmission of x-rays through tissue, and the spatial variability of transmission is measured to create a visualization of internal anatomy. According to physics, the transmission of photons is a consequence of mainly two photon interaction phenomena: absorption and scatter. Specifically, at diagnostic clinical x-ray energies, transmission photons interact with tissue via photoelectric, Compton and Rayleigh processes, characterized by the associated linear attenuation properties correlating with interaction type, energy, and tissue. The collective interactions divert the source photons from a primary beam which can be measured and used to define total attenuation. The scattering process is largely considered destructive to imaging information content, and radiological innovation has historically improved through enhanced scatter discrimination to better discard the unwanted, scattered photons [1, 2].

Compton scattering is the dominant process for biological (water equivalent) materials in clinical x-ray systems energies (50-500 keV photon energies. It is a characterizable phenomenon and produces patterns that are unique to an object's density (electron density) and its location in an x-ray beam. While x-ray attenuation imaging, based on x-ray absorption, is straightforward and effective, it has been recognized in numerous scientific papers, that scattered photons contain additional information [3-8]. This sentiment was well articulated by Greenberg et. al, in discussing the topic: "…each photon carries a wave vector, polarization, and energy associated with the location and type of scatter. These extra dimensions of information provide additional contrast mechanisms, which can lead to novel diagnostic capabilities" [4]. Indeed, in the past two decades, and particularly in recent years, we have seen novel approaches for harnessing enhanced information in photon transmission imaging. Proof of principle papers have been published in the fields of crystalography, coherent scatter tomography, coding and sampling, and scatter projection [4, 9-11]. Existing and emerging literature demonstrates multiple approaches for utilizing scatter information and warrants continued investigation. Perhaps most commonly studied are techniques based on small angle coherent scatter [12-14] – these techniques utilize principles of Reyleigh scattering and require lower energy x-rays, often high x-ray fluence, and physically smaller imaging samples, and likely cannot scale to general in-vivo clinical x-ray utility. While utilization of higher energy compton scatter has been considered, solutions have been limited and analytical constructions – i.e. fully mathematically modelled systems [7, 10, 15-17].

As we approach the challenge of scatter utilization, it is prudent to recognize that recent years have seen computing power increasing exponentially. Also, the field of artificial intelligence (AI), specifically the subfield of deep learning (DL), has matured dramatically, making it more powerful and more accessible up to the point that it is ready to be fully integrated into diverse applications. Many fields, including medical imaging, are now in the process of adapting to this paradigm shift. The introduction of DL, and specifically convolutional neural networks (CNNs) allow us to harness information rich data to extract additional features. An opportunity in this new DL driven imaging paradigm then comes from revisiting the fundamental design of current x-ray imaging systems and optimize them for information capture. In initiating this work, we hypothesized an x-ray system setup where we can extract target unique scattering patterns to be used as input for training a CNN. Furthermore, contemporary detectors can provide these scattering patterns in a multitude of energy bins, providing an additional layer of information gain to support new techniques [10, 18].

In this work, we propose a keV x-ray imaging system design that improves upon current 2D x-ray imaging systems through measurement and use of the Compton scattering effect. Using modern hardware and software techniques, we strive to extract a new spatial dimension – material class/electron density along the beam. This information is derived from off-axis scattered photons and DL supported depth reconstruction. The presented work is limited to extract material classes in a controlled system design. We built a virtual planar x-ray imaging system and tested it on geometrically distributed body tissues via Monte Carlo (MC) simulations. Our ability to discern material along the projection profile is reported. This work is limited to elementary simulations and intended to provide preliminary proof of principle and support further development towards clinically practical systems.

## 2. Materials and Methods

### 2.1 Apparatus

The setup we propose is based on a pencil beam probing system, as depicted in Figure 1. This system is inspired by previously proposed x-ray system designs [6, 12, 19] built with similar ambitions, however with the significant modifications (i) that it does not use scatter grid/coded aperture thus making use of all available photons, and (ii) the signal recovery (resolution of superimposed anatomy) in the system is driven by DL. The system takes the classic design of a standard x-ray transmission setup but collapses the x-ray source to a pencil beam which allows for the Compton scatter information along the beam axis to be isolated across the projection plane. By doing this we optimize the information gathered from each photon traversing the object, scattered and unscattered, and without imposing any increase in radiation dose.

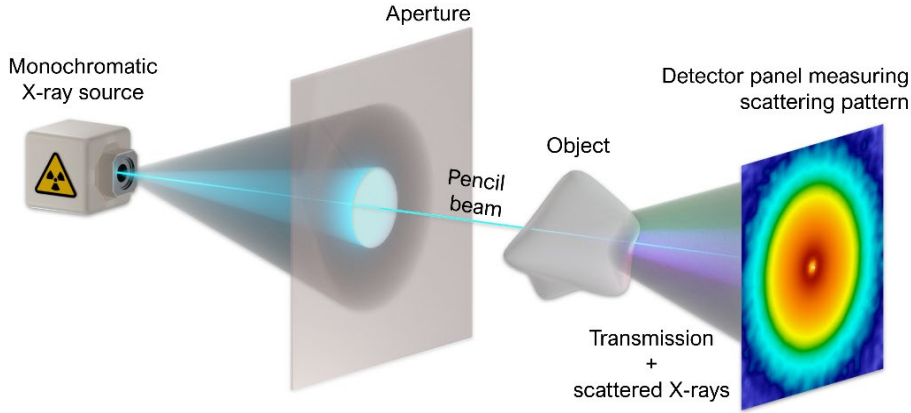

**Figure 1.** Schematic of an envisioned pencil beam scatter reconstruction system. A monochromatic x-ray beam is focused to a narrow beam to be incident on an object. The subsequent photon transmission and scatter pattern is captured by a 2D planar detector panel. The envisioned system then uses DL to process the detected patterns to resolve traversed depth information. In this study, our simulation uses a pure pencil beam source.

We modelled a typical photon beam using MC ray tracing with accurate scatter representation. These scattered and transmitted photons make up the signal that is detected across the face of the detector and across the detector's energy bins, as seen in Figure 2. Detected photon fluence exhibit concentric ring patterns on the face of the detector, specific to energy, and becoming more asymmetric with bigger and more diverse geometries. This detected signal, apportioned into energy bins, form the input for our proposed CNN-based reconstruction algorithm. Ultimately, we strive to acquire a 1D traversed tissue map for each probing pencil beam. Once a 1D profile capture system is created, it could potentially be extended to a 3D imaging system by raster scanning, and/or possibly extended to fan beams [8].

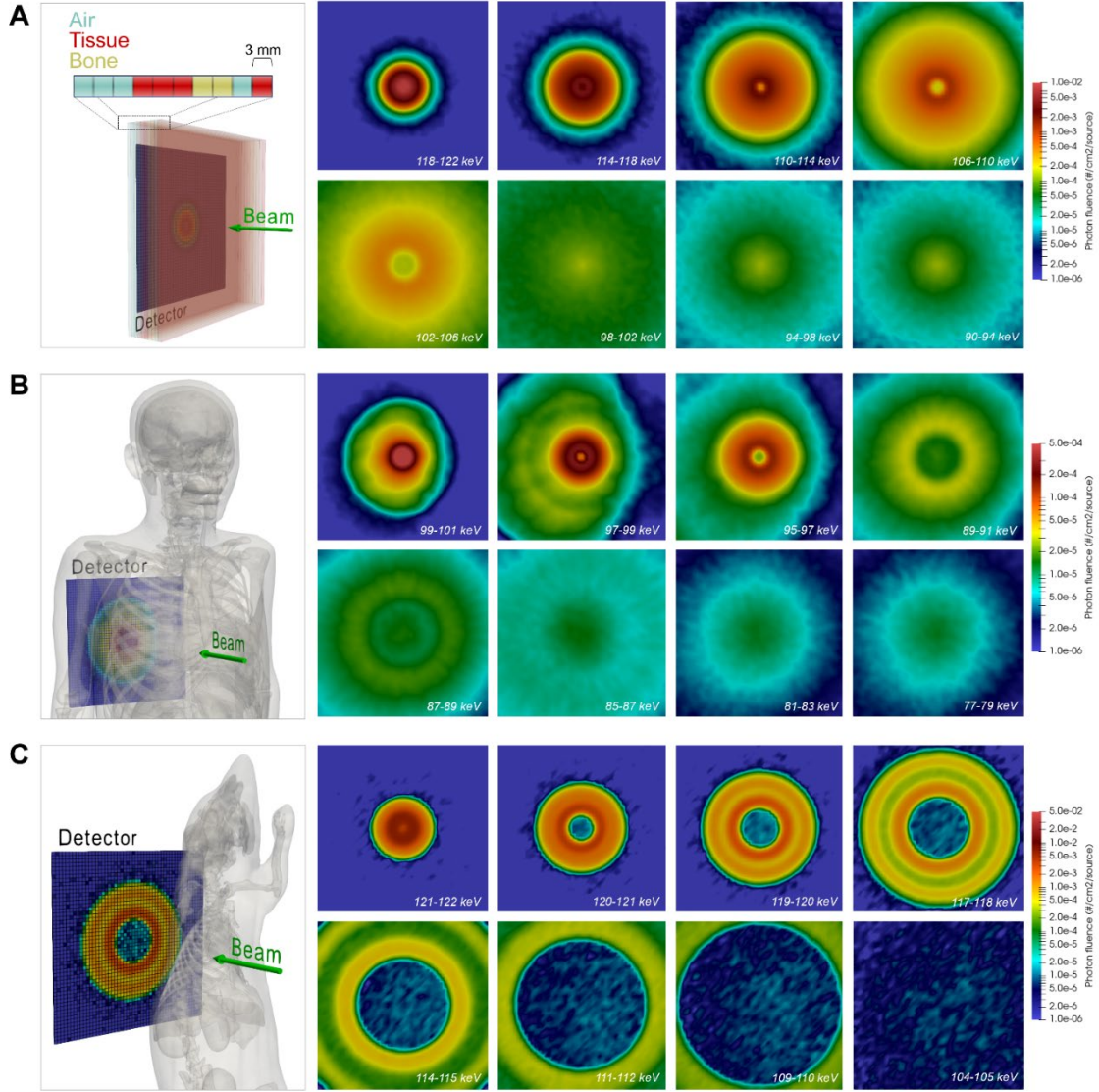

**Figure 2.** Simulation of detector readout per instance/phantom/beam. Detected photon fluence patterns are shown for a spectrum of energy windows for different setups: (A) Project simulation - a stack of tissue slabs, an example of the setup used in this study, with an incident beam of 122 keV, (B, C) Envisioned applications with MC simulated spectra – (B) an incident beam of 100 keV on a simulated human, and (C) an incident beam of 122 keV on a simulated mouse. Only the slabs setup shown in (A) was studied in this project, but the other scenarios were simulated and presented to demonstrate there is opportunity to extend our work to more complex scenarios as they produce similar energy-spatial scatter patterns.

## 2.2 Experimental simulation

We trained and tested our imaging apparatus using a simple simulation where the imaging object consisted of ten stacked slabs of differing tissue classification (Figure 2A). Each slab was 3 mm thick and comprised air ($\rho=0.0012$ g/cm3), soft tissue ≈ water ($\rho=1.0$ g/cm3), or bone ($\rho=1.85$ g/cm3). The assigned material was pseudo-random, with conditions that tissue materials would inhibit between two and five adjacent slabs (to mimic biological samples). A monochromatic 122 keV photon beam of infinitesimal thickness was oriented normal to the slab phantom 3.0 mm above the phantom surface (i.e. passing through 3.0 mm of air prior to reaching the phantom). The energy of 122 keV was chosen as it corresponds to the principal emission of Co-57, a relatively accessible isotope, representing one option for beam generation. At the opposing side of the phantom, a 51×51 element planar photon fluence detector (2.4 mm isotropic element size) was oriented 3.0 mm from the phantom face, and normal to the beam source. Photon fluence through the detector was scored as a function of energy, using small 1.0 keV energy bins from 30-130 keV. These small bins allowed us to group and study different energy resolution scenarios.

## 2.3 Monte Carlo simulation

376 phantom imaging studies were generated via MC. A total of $2.5\times10^8$ histories (photons) were simulated for each phantom using the Particle and Heavy Ion Transport code System (PHITS v3.16). All simulations utilized the PHITS-EGS5 method for treatment of multiple scattering, explicit treatment of fluorescent x-rays, consideration of Rayleigh and incoherent scattering, and consideration of electron-impact ionization. The photon cutoff energy was set to 1.0 keV.

## 2.4 DL model design and training procedure

The 376 phantoms were randomly split into two sets: 300 were used for model training and tuning, and 76 for held-out testing. To investigate effects of the random data split, the training set was randomly split 30 times into 240 training and 60 tuning cases, and each split was trained and tested independently.

We used a multiclass deep neural network to infer the 10 traversed pixels based on the measured scatter pattern. An overview of the DL model is shown in Figure 3. The ResNet50 CNN [19] was used to extract features from the 2D energy images, followed by a fully connected layer of output size 30 (10x3) to represent a one-hot encoding of the three possible material classes (air, tissue and bone) of the 10 traversed pixels. Eight different input data energy resolutions were investigated by combining the 1 keV energy bins, ranging from 1 to 100 keV (100 to 1 input channels) and the input image stacks were z-normalized. The multiclass model was trained with a batch size of 61 over 1000 epochs using a cross entropy loss function. Four random initializations were run for each setting. Both stochastic gradient descent and the adam optimizer [20] was tested, and the learning rate was optimized by a simple grid search. Furthermore, three different augmentation schemes were tested: random horizontal, vertical and combined horizontal and vertical flipping of the image stack. The model of the last epoch of the best initialization was used for subsequent testing. The tuning set accuracy was used to compare settings. The final model used the adam optimizer with learning rate=1e-6. Neither augmentation scheme improved results and were hence omitted. The model was implemented in PyTorch (v1.4) and trained on Nvidia GTX 1080/1080Ti GPUs.

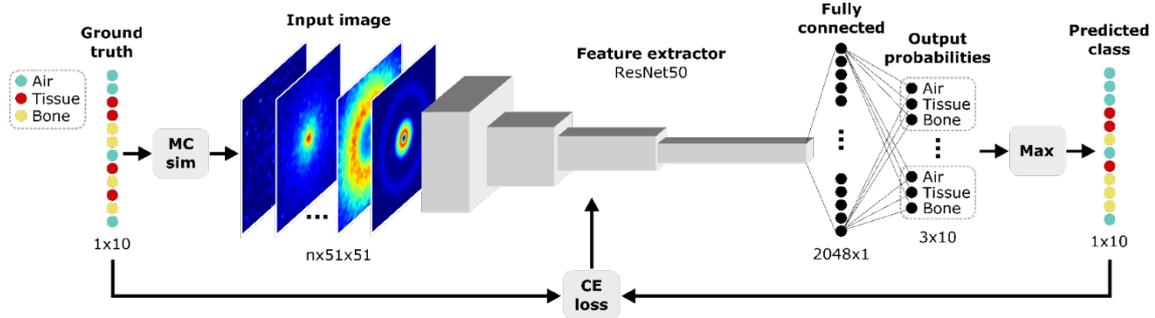

**Figure 3.** Schematic view of the multiclass deep neural network model. The ground truth depth pixels of different materials were MC simulated to obtain realistic scatter images, which were sorted into n energy bins and used as input to a ResNet50 model. The 2048 features from ResNet50 were fully connected to 30 (3x10) output nodes, and the class with maximum probability was assigned to each of the 10 nodes. The model was updated using backpropagation of the cross entropy (CE) loss.

## 3. Results

The average classification accuracy of the 10 depth pixels for the held-out test set (n=76) run with the 30 split models, using an energy resolution of 4 keV was 0.91±0.01, seen in Figure 4A. The average sensitivity (recall) and specificity was 0.91 and 0.95 respectively. The confusion matrices for precision and recall for the three materials are seen in Figure 4B and C. Air is best classified with a 96% accuracy, followed by bone at 91% and tissue at 86%. As expected, the largest confusion happens between tissue and bone, which both have a higher, more similar density compared to air. Figure 5 shows the average accuracy per position of the 10 traversed pixels. The model generally performs better for pixels closer to the detector (towards the right), possibly due to the enhanced likelihood of signal from single scatter event photons. Pixels closest to the bean entrance also exhibited higher accuracy than internal pixels, possibly indicating an advantage of a higher fluence traversing the pixel thus providing more signal. The classification performance relative the energy bin width for the tuning set is depicted in Figure 6. A width of 1-4 keV (25-100 channels) yields similar accuracy and starts dropping at >5 keV (20 channels). Here we are using a width of 4 keV which is aligned with state-of-the art detector resolutions [18, 21].

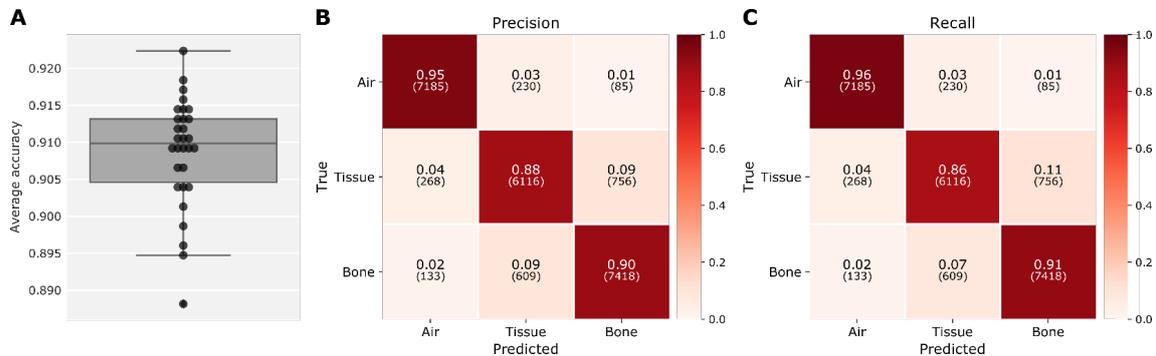

**Figure 4.** Classification performance on the held-out test set (n=76) for the network with 4 keV energy resolution. Shown are the (A) average accuracy of the 10 depth pixels for each of the 30 best (of 4 initializations) models, and the confusion matrices of (B) precision, and (C) recall, showing >0.88 precision and >0.86 recall (sensitivity) for all three materials.

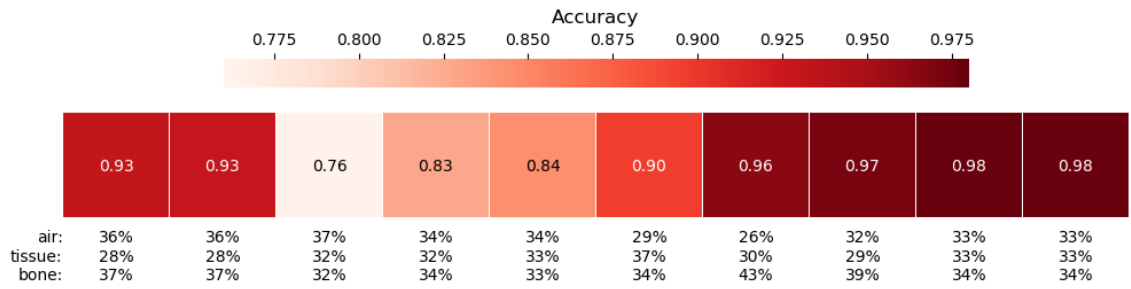

**Figure 5.** The average accuracy of the test set (n=76) for each of the 10 traversed pixels from the source to the detector (left to right), and the occupancy fraction of the three material classes for those pixels.

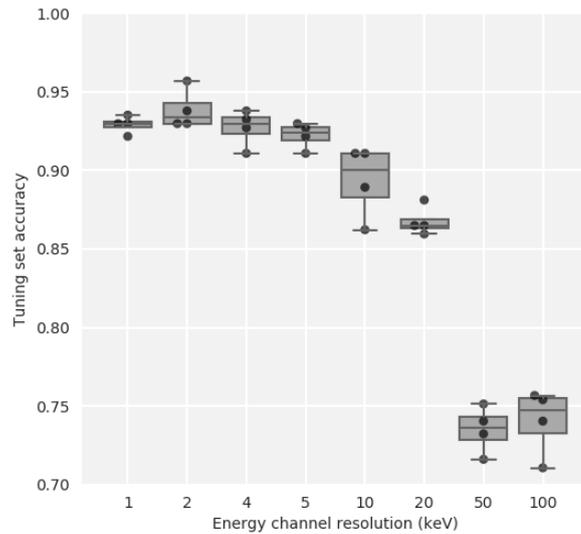

**Figure 6.** Tuning set (n = 37) accuracy for one split for an input channel energy resolution of 1, 2, 4, 5, 10, 20, 50, and 100 keV, corresponding to 100, 50, 25, 20, 10, 5, 2, and 1 input channels. The training for each setting was randomly initialized four times, seen as the four markers of each box. In this study, we focus on a 4 keV energy resolution..

## 4. Discussion

In this work we showed that with modifications to a traditional x ray source-object-panel detector framework, specifically optimizing beam shape, energy discrimination and event detection localization, and using high energy resolution panel detectors, we can resolve traversed depth material via DL aided reconstruction, without increasing radiation dose.

This work provides proof of principle showing that Compton scattered photons can be used to inform about the tissue they traverse. This demonstrated a principle that the information that exists in the signal can be extracted with CNNs. However, this work is limited to proof of principle, using technological specifications we project for in current or coming years – specifically for the purpose of pioneering relevant technology for the future. The study design was simplistic and almost every element of the setup and DL architecture needs to be further tuned and investigated. These items include: Pencil beam generation and its inclusion of penumbra, not modelled here. The optimal energy and energy spectrum of photon beams also needs to be studied; we modelled a 122 KeV beam because this can simply be generated with the Co-57 isotope. The most obvious/easy source of monochromatic photos would be from synchrotrons, but those are not economically practical for robust use. Excitingly, alternative monochromatic x-ray sources are being developed that may also be considered [22]. Furthermore, the use of a bremsstrahlung photon spectrum rather than monoenergetic photons can be investigated. Such a system would be simpler to design as it resembles current x-ray imaging technology, although the signal would be more complex and one would likely require significantly more data to train the DL model. The size, shape and performance of real detector elements need to be further considered. The setup, design, and implementation of the core CNN DL reconstruction model also needs to be examined, tuned and trained for more robust application. This area may have significant opportunity supported by potentially very large training sets (clinical CTs can be used) and expanded CNN models, which for example might include adjacent beam information.

The ability to generate complex images that contain both attenuation and depth information could, if successfully developed, represent a leap forward in x-ray technology. It could enable a dimensional expansion on contemporary planar x-ray technology and introduce 3D imaging into areas where it is not presently practical or possible. It could provide clinicians with more detailed/information-rich images from which to provide diagnosis, and industry with a new capacity for innovation. It could reduce radiation dose from more efficient use of signal and collimator removal.

In this proof of principle work, we designed our DL model to solve a multiclass classification task, namely classifying each traversed pixel into one of three materials. The model can be redesigned and/or expanded to solve a regression task, where instead of classifying each pixel, their Hounsfield value or such can be inferred. Instead of minimizing the cross entropy, for such task one can minimize the mean squared error or similar. Because this is a more complicated task, more training data would likely be required. Elegant to this design, complex and robust CNN training could likely be achieved with ubiquitously available CT data, which can provide both 2D and 3D MC simulation models and gold standard "truth". Especially for simple designs, care has to be taken to ensure training and test data are not too similar, inflating the performance. Here the average similarity (fraction of the 10 pixels positions having identical material) between training and test pairs was 33%.

The redesign of an x-ray system we are proposing was focused on adding a spatial dimension along the beam axis. However, our system also expands the information content per photon available in 2D x-ray setup and can likely form the basis for other DL assisted x-ray imaging improvements. If/when physical imaging systems supply information rich data like we are proposing, we will likely see many directions of innovation around its use; for example, in functional k-edge tracer imaging [23].

## 5. Conclusion

While limited by a simplified experimental design, the results of this feasibility study demonstrate: (i) Compton scatter contains information that can be used to resolve superimposed anatomy, and (ii) our DL model was capable of identifying the traversed materials with high accuracy. Future efforts are still needed to characterize the capacities and limitations of this technique. As advances in hardware and DL architectures are introduced to the field, we can expect Compton scatter utilization to be more and more practical, making this an interesting area for investigation.

## Acknowledgments

The authors are grateful for the generous computational support given by the Warren Alpert foundation. This research was funded in part through the NIH/NCI Cancer Center Support Grant [grant number P30 CA008748].